\documentclass[preprint,groupedaddress,showpacs]{revtex4}
\usepackage{amssymb}
\usepackage{amsmath}
\usepackage{dcolumn}
\usepackage{bm}
\usepackage{graphicx}
\usepackage{mathrsfs}
\usepackage{appendix}

\begin{document}

\title{Tunable electromagnetically induced transparency with a coupled superconducting system}
\author{Xin Wang, Hong-rong Li$^{\dag}$, Wen-xiao Liu, and Fu-li Li}
\address{Institute of Quantum Optics and Quantum Informations; School of Science, Xian Jiaotong University, Xian 710049, China}

\begin{abstract}
Electromagnetically induced transparency (EIT) has usually been demonstrated by using three-level atomic systems. In this paper, we theoretically
proposed an efficient method to realize EIT in microwave regime through a coupled system consisting of a flux qubit and a superconducting $LC$
resonator with relatively high quality factor. In the present composed system, the working levels are the dressed states of a two-level flux qubit and the
resonators with a probe pump field. There exits a second order coherent transfer between the dressed states. By comparing the results with those in
the conventional atomic system we have revealed the physical origin of the EIT phenomenon in this composed system. Since the whole system is artificial and tunable, our scheme may have potential applications in various domains.\newline
$^{\dag}$Corresponding Email: hrli@mail.xjtu.edu.cn\newline
\end{abstract}

\maketitle

\ \ \ \ \ \ \ \ \ \ \ \ \ \

%\keywords{}

\section{Introduction}

The electromagnetically induced transparency (EIT), which manifests spectroscopically the quantized three-level structures of an atomic medium
through its interaction with two semiclassical fields, was first observed in atomic gases \cite{[1]}. As a powerful technique that can be used to
eliminate the effect of a medium on a propagating beam of electromagnetic radiation, EIT has been widely explored in various aspects \cite{[2],[3],[4]}. It is also of great application potential in non-linear optics and quantum information \cite{[5]}, for example, using the EIT-based slow and even store
light to realize controllable optical delay lines and optical buffers \cite{[6]}. In recent years schemes realizing EIT by using artificial systems have
drawn much attention \cite{[7]}. Weis \textit{et al.} \cite{[8]} demonstrated that there exists optomechanical induced transparency and they have found some similarities with the conventional EIT in the atomic systems. In Ref. \cite{[9]}, the authors also proposed an EIT scheme in an asymmetric double quantum dot system. Compared with that of the atomic systems, EIT based on the artificial systems is more adjustable and thus has more potential applications.

Recent development of superconducting quantum devices provides another vital artificial quantum system. Because of more easily designed and fabricated on demand, superconducting quantum circuits have been used to test many quantum optical phenomena \cite{[10],[11]}. As one of the basic types of superconducting qubits, the flux qubit is often treated as a two-level system and has been theoretically and experimentally investigated in the
fields of quantum information and quantum optics \cite{[12]}. In this work, we will show the possibilities to obtain EIT by a flux qubit coupled to a
superconducting $LC$ resonator. The coupling between the qubit and the resonator is of diagonal form ($\sigma _{z}$-coupling). In this system,
coupling to the $LC$ resonator provides the flux qubit with additional energy levels to realize EIT phenomena.

Our proposal can also work as the efficient optical devices controlling slow light and fast light in microwave regime as shown in Refs. \cite{[13],[14],[15]} by adjusting pump field, which has been discussed in detail in those references. Here we will only study the physical basis of EIT in our proposal, and by employing the effective Hamiltonian methods we find the second order coherent transfer between the dressed states plays an important role in EIT of this coupling type ($\sigma _{z}$-coupling) system for the first time. Comparing our results with those of conventional atomic EIT, we point some similarities and differences between them. In addition, our scheme is more adjustable than that of the atomic systems and may have potential applications on microwave photonics, nonlinear optics and optical communication in microwave regime \cite{[13]}. Moreover, the whole system is based on all superconducting quantum circuits, so it can be integrated on a chip and has a small size.

The organization of this paper is as follow: We introduce our model in section 2. The analytical deductions and discussion of EIT are shown in section 3. The conclusion is made in section 4.

\section{Model}

Our proposal is illustrated in Fig. 1. The quantum device consists of a superconducting flux qubit \cite{[16],[17]} and a superconducting $LC$
resonator made of a capacitor $C$ and an inductor $L$ \cite{[18],[19]}. As demonstrated in Ref. \cite{[20]}, the Hamiltonian of the gap-tunable flux
qubit composed of a qubit loop and a superconducting quantum interference device (SQUID) can be expressed as $H_{q}=[\frac{1}{2}\epsilon (\Phi _{z})\bar{\sigma}_{z}+\Omega (\Phi _{x})\bar{\sigma}_{x}]$ in the basis of persistent current states \{$|+\rangle ,|-\rangle $\}, where $\bar{\sigma}_{z}$ $=|+\rangle \langle +|-|-\rangle \langle -|$ and $\bar{\sigma}_{x}=|+\rangle \langle -|+|-\rangle \langle +|$ are the Pauli operators in the basis of clockwise $|+\rangle $ and anticlockwise $|-\rangle$ persistent currents. $\epsilon $ and $\Omega $ can be controlled via the external flux $\Phi _{z}$ and $\Phi _{x}$ independently.

For the first term of $H_{q}$, $\epsilon (\Phi _{z})=2I_{p}(\Phi _{z}-\Phi_{0}/2)$ is the energy bias between the clockwise $|+\rangle $ and anticlockwise $|-\rangle $ of persistent current $I_{p}$ in the qubit loop, $\Phi _{0}=\hbar /2e$ is the flux quantum, and $\Phi _{z}$ is the external flux that can be induced by flux driving through the qubit loop via a microwave (MW) current line. The flux driving can be separated into a static part $\phi $ and a time-dependent part $\mu _{0}SI(t)/2\pi l$, where $\mu_{0}$ is the vacuum permeability, $l$ is the distance between the MW line and the qubit loop, $I(t)$ is the MW current, and $S$ is the effective coupling area of the qubit loop for the current line \cite{[13]}. We set $\phi =\Phi _{0}/2$ to minimize the flux noise. As a result, the first term of $H_{q}$ includes a driven term and can be expressed as $H_{d}=\mu I(t)\bar{\sigma}_{z}$, where $\mu =\frac{\mu _{0}SI_{p}}{\pi l}$ is the effective "electric dipole moment" of the qubit \cite{[13]}.
\begin{figure}[tbph]
\centering\includegraphics[width=7cm]{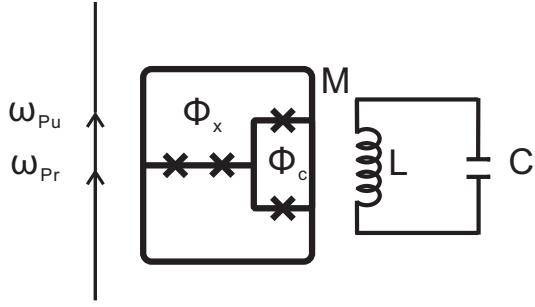}
\caption{Schematic of the composed system between flux qubit and superconducting LC resonator. The microwave currents with $\protect\omega_{pu}$ and $\protect\omega _{pr}$ are applied through the qubit loop.}
\label{fig1}
\end{figure}

For the second term of of $H_{q}$, $\Omega (\Phi _{x})$ is the qubit gap and depends on flux driving $\Phi _{x}$ of the SQUID loop. Here we consider the SQUID loop is coupled with a superconducting $LC$ micrometer resonator \cite{[21]} that can be described by a simple harmonic oscillator Hamiltonian $H_{R}=\hbar \omega (b^{\dagger }b+1/2)$, where $b^{\dagger }(b)$ is the creation (annihilation) operator for the resonator. The resonance frequency of the resonator is given by $\omega =1/\sqrt{LC}$, and the value of the frequency is in the range of hundreds of MHz to several GHz with a quality factor $Q\sim $10$^{3}$-10$^{6}$ \cite{[12]}. The $LC$ resonator interacts with the SQUID loop of the flux qubit via the mutual inductance. With a static part $\hbar \upsilon /2$ we can express the qubit gap as $\Omega (\Phi _{x})=$ $\hbar \upsilon /2+\hbar \frac{\partial \Omega (\Phi _{x})}{\partial \Phi _{x}}|_{\Omega (\Phi _{x})=\hbar \upsilon /2}(M\sqrt{\omega /2\hbar L})(b^{\dagger }+b)$ \cite{[18],[21]}, where $\frac{\partial \Omega (\Phi _{x})}{\partial \Phi _{x}}|_{\Omega (\Phi _{x})=\hbar \upsilon /2}$ is the sensitivity of the SQUID loop to the flux \cite{[22]}, and $M$ is the mutual inductance between the SQUID loop and the resonator, and $\sqrt{\hbar \omega /2L}$ is the amplitude of the vacuum fluctuation of the current in the $LC$ oscillator. Supposing $g=\frac{\partial \Omega (\Phi _{x})}{\partial \Phi _{x}}|_{\Omega (\Phi _{x})=\hbar \upsilon /2}(M\sqrt{\omega /2\hbar L})$, we rewrite $H_{q}$ as
\begin{equation}
H_{q}=\mu I(t)\bar{\sigma}_{z}+\frac{1}{2}\hbar \upsilon \bar{\sigma}_{x}+\hbar g(b^{\dagger }+b)\bar{\sigma}_{x}.
\end{equation}

In the new basis of the eigenstates of the qubit \{$|e\rangle = \{(|+\rangle +|-\rangle )/\sqrt{2}$, $|g\rangle =(|+\rangle -|-\rangle )/\sqrt{2}\}$, the Hamiltonian of the system can be expressed as
\begin{equation}
H_{sys}=H_{q}+H_{R}=\frac{1}{2}\hbar \upsilon \sigma _{z}+\hbar \omega b^{\dagger }b+\hbar g(b^{\dagger }+b)\sigma _{z}+\mu I(t)(\sigma
^{+}+\sigma^{-}),
\end{equation}
where $\sigma _{z}=|e\rangle \langle e|-|g\rangle \langle g|,\sigma^{+}=|e\rangle \langle g|,\sigma ^{-}=|g\rangle \langle e|,$[$\sigma_{z},\sigma ^{\pm }$]=$\pm 2\sigma ^{\pm },$[$\sigma ^{+},\sigma ^{-}$]=$\sigma _{z}$. We consider two driving currents: pump and probe currents with amplitude $\varepsilon _{pu}$ and $\varepsilon _{pr}$ at frequency $\omega _{pu}$ and $\omega _{pr}$ respectively; that is, $I(t)=-2[\cos(\omega_{pu}t)\varepsilon _{pu}+\cos (\omega _{pr}t)\varepsilon _{pr}]$ ($\varepsilon _{pu}$ $\gg $ $\varepsilon_{pr})$. By applying a frame rotating at frequency $\omega _{pu}$ and adopting the rotating wave approximation, the Hamiltonian of the system reduces to (we set $\hbar =1)$

\begin{equation}
H_{sys}=\frac{1}{2}\Delta \sigma _{z}+\omega b^{\dagger }b+g(b^{\dagger}+b)\sigma _{z}-\Omega _{pu}(\sigma _{+}+\sigma _{-})-\Omega _{pr}(\sigma
_{+}e^{-i\delta t}+\sigma _{-}e^{-i\delta t}),
\end{equation}
where $\Delta =\upsilon -\omega _{pu}$ is the pump-exciton detuning, $\Omega_{pu}=\frac{\mu \varepsilon _{pu}}{\hbar }$ $(\Omega _{pr}=\frac{\mu
\varepsilon _{pr}}{\hbar })$ is the Rabi frequency of pump current (probe current), and $\delta =\omega _{pr}-\omega _{pu}$ is the signal-pump detuning.

\section{Numerical results and theoretical analysis}

In this section we will employ the semiclassical approach to explore the EIT phenomenon in our proposal. Taking the semiclassical approach, we can derive the equations of motion for $\sigma _{z}$, $\sigma _{-}$, and $X=b^{\dagger}+b$ by applying the Heisenberg equations and introducing the
corresponding damping terms. The equations of motion for each operator read as follows \cite{[21]}:
\begin{equation}
\frac{d\sigma _{z}}{dt}=-(\sigma _{z}+1)\Gamma _{d}+2i\Omega_{pu}(\sigma_{+}-\sigma _{-})+2i\Omega _{pr}(\sigma _{+}e^{-i\delta t}-\sigma_{-}e^{i\delta t}),
\end{equation}

\begin{equation}
\frac{d\sigma ^{-}}{dt}=[-\Gamma _{f}-i(\Delta +2gX)]\sigma^{-}-i\Omega_{pu}\sigma _{z}-i\Omega _{pr}\sigma _{z}e^{-i\delta t},
\end{equation}

\begin{equation}
\frac{d^{2}X}{dt}+\gamma \frac{dX}{dt}+\omega ^{2}X=-2\omega g\sigma _{z}.
\end{equation}

In the above equations, $\Gamma _{d}$ and $\Gamma _{f}$ are the energy decay rate and dephasing rate for the flux qubit respectively, and $\gamma =\omega /Q$ is the decay rate for the $LC$ resonator coupling to a reservoir of "background" modes and other intrinsic processes. We have taken the semiclassical approach that ignores the correlation between the qubit and the resonator, i.e. $\langle X\sigma _{z}\rangle =\langle X\rangle \langle
\sigma _{z}\rangle $. In order solve the equations above we make the ansatz $X=\bar{X}_{0}+\langle X\rangle ^{+}e^{-i\delta t}+\langle X\rangle
^{-}e^{i\delta t}$, $\sigma _{-}=\bar{\sigma}_{-}^{0}+\langle \sigma_{-}\rangle ^{+}e^{-i\delta t}+\langle \sigma _{-}\rangle ^{-}e^{i\delta t}$, and $\sigma _{z}=\bar{\sigma}_{z}^{0}+\langle \sigma _{z}\rangle^{+}e^{-i\delta t}+\langle \sigma _{z}\rangle ^{-}e^{i\delta t}$. Since the pump current is of much stronger than the probe (signal) current, we work to all orders in $\varepsilon _{pu}$ but to the lowest order in $\varepsilon_{pr}$, then we can obtain $\langle \sigma _{-}\rangle ^{+}$, which corresponds to the dimensionless effective linear susceptibility as follows \cite{[13],[15]}: $\chi _{s}^{1}(\omega _{pr})=\frac{\mu \langle \sigma _{-}\rangle ^{+}}{\varepsilon _{pr}}=\frac{\mu ^{2}}{\hbar \Gamma _{f}}\chi ^{1}(\omega _{pr}),$ and $\chi ^{1}(\omega _{pr})$ is given by

\begin{equation}
\chi ^{1}(\omega _{pr})=i\Gamma _{f}\frac{(\Omega _{pu}+2g\bar{\sigma}_{-}^{0}C)\{\frac{2\Omega _{pu}\bar{\sigma}_{z}^{0}A^{-1}+2i(\bar{\sigma}_{-}^{0})^{\ast }}{\Gamma _{d}-i\delta -2i\Omega_{pu}[i\Omega_{pu}(B^{-1}-A^{-1})+2igC(B^{-1}(\bar{\sigma}_{-}^{0})^{\ast}-A^{-1}\bar{\sigma}_{-}^{0})]}\}+\bar{\sigma}_{z}^{0}}{A},
\end{equation}
where $A=i\delta -\Gamma _{f}-i\Delta -2ig\bar{X}_{0}, B=\Gamma_{f}-i\delta-i\Delta -2ig\bar{X}_{0}, C=\frac{-2\omega g}{\omega ^{2}-\delta^{2}-i\gamma \delta }$, and the population inversion $\bar{\sigma}_{z}^{0}$ is the solution of the equation below:
\begin{equation}
\Gamma _{d}\bar{\sigma}_{z}^{0}[(\Delta -4g^{2}\bar{\sigma}_{z}^{0}/\delta)^{2}+\Gamma _{f}^{2}]+4\Gamma _{f}\Omega _{pu}^{2}\bar{\sigma}_{z}^{0}+\Gamma _{d}[(\Delta -4g^{2}\bar{\sigma}_{z}^{0}/\delta)^{2}+\Gamma _{f}^{2}]=0,
\end{equation}
where $\bar{X}_{0}$ and $\bar{\sigma}_{-}^{0}$ can be obtained via the relations $\bar{X}_{0}=\frac{-2g\bar{\sigma}_{z}^{0}}{\omega}$ and $\bar{\sigma}_{-}^{0}=\frac{-\Omega _{pu}\bar{\sigma}_{z}^{0}}{(\Delta +2g\bar{X}_{0})-i\Gamma _{f}}$.

For illustration of the numerical results, we choose some reasonable parameters. The resonant frequency of the superconducting $LC$ resonator is $\omega =1$GHz \cite{[19]} with quality factor $Q=10^{4}$ (corresponding $\gamma =0.1$MHz) \cite{[12]}, and the coupling strength can be $g=80$MHz \cite{[21]}. The decoherent rates of the qubit $\Gamma _{d}=60$MHz and $\Gamma _{f}=30$MHz \cite{[23]}. We suppose the driving field strength $\Omega _{pu}=50$MHz with pump-exciton detuning $\Delta=\omega =1$GHz. In Fig. 2 we plot the real and image part of $\chi^{1}(\omega _{pr})$, respectively.

\begin{figure}[tbph]
\centering\includegraphics[width=8cm]{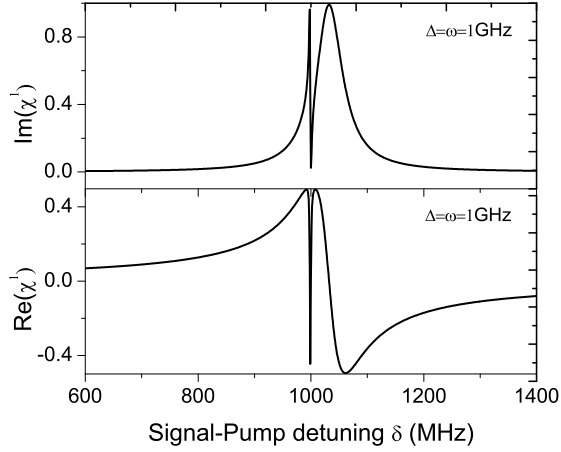}
\caption{The absorption$\ $Im$\protect\chi ^{1}(\protect\omega _{pr})$ and dispersion Re$\protect\chi ^{1}(\protect\omega _{pr})$ of the probe current
as a function of the signal-pump detuning $\protect\delta$ under the condition $\Delta =\protect\omega =1$GHz.}
\label{fig2}
\end{figure}

Fig. 2 illustrates the behavior of the absorption Im$\chi ^{1}(\omega _{pr})$ and dispersion Re$\chi ^{1}(\omega _{pr})$ of the probe current as a
function of signal-pump detuning $\delta =\omega _{pr}-\omega _{pu}$. We find that at $\delta =\omega $ the absorption Im$\chi ^{1}(\omega _{pr})$ is near zero with a steep positive slope of dispersion Re$\chi ^{1}(\omega _{pr})$ (corresponding to the slow light effects \cite{[13]}), which mean EIT
happens around this regime.

In the following, we will obtain a more simple form of $\chi ^{1}(\omega_{pr})$ and discuss how EIT happens in this artificial system. With no
consideration of the weak signal current, the Hamiltonian of the system in Eq. (3) is then
\begin{equation}
H_{sys}^{^{\prime }}=\frac{1}{2}\Delta \sigma _{z}+\omega b^{\dagger}b+g(b^{\dagger}+b)\sigma _{z}-\Omega _{pu}(\sigma _{+}+\sigma_{-}).
\end{equation}

Performing a unitary transformation $U(t)=$ $\exp (iH_{0}t)$ with $H_{0}=\frac{1}{2}\Delta \sigma _{z}+\omega b^{\dagger }b$ to $H_{sys}^{^{\prime }}$, and under the condition $\Delta =\omega $ we will obtain the effective Hamiltonian of the system as below \cite{[24]}:

\begin{equation}
H_{eff}=\frac{\Omega _{pu}^{2}}{\Delta }\sigma _{z}+\frac{2g\Omega _{pu}}{\Delta }(b^{\dagger }\sigma _{-}+b\sigma _{+}),
\end{equation}
where the first term of $H_{eff}$ is the energy shift induced by the driving current. As shown in Fig. 4, the second term describes a coherent transfer
between states $|g,n+1\rangle $ and $|e,n\rangle $ ($|n\rangle $ being the Fock state of the $LC$ resonator), it corresponds to a second order
coupling between these two states with strength $\frac{2g\Omega _{pu}}{\Delta }$.

To verify $H_{eff}$ and $H_{sys}^{^{\prime }}$ are consistent, with assuming that the initial state of system is $|g,1\rangle $, we define the population
possibilities $P_{e0}(t)=Tr[|e,0\rangle \langle e,0|\rho (t)]$, $P_{g1}(t)=Tr[|g,1\rangle \langle g,1|\rho (t)]$ and $P_{g0}(t)=Tr[|g,0 \rangle \langle g,0|\rho (t)]$, where $\rho (t)$ is the density matrix of the system. We then proceed to numerically solve the master equations with $H_{i}=H_{sys}^{^{\prime }}$ and $H_{i}=H_{eff}$ respectively for the whole system, which can be written as
\begin{equation}
\frac{d\hat{\rho}(t)}{dt}=-i\left[ H_{i},\hat{\rho}(t)\right] +D[\sigma^{-},\Gamma _{f}]\hat{\rho}(t)+D[b,\gamma ]\hat{\rho}(t),
\end{equation}
where $D[A,\Omega ]\hat{\rho}=\Omega (A\hat{\rho}A^{+}-\frac{1}{2}A^{+}A\hat{\rho}-\frac{1}{2}\hat{\rho}A^{+}A)$ are the decoherent terms of Lindblad
form.

In Fig. 3 we display the numerical results for time evolution of population possibilities. We find that evolutions determined by $H_{eff}$ and $H_{sys}^{^{\prime }}$ match well and a coherent transfer between state $|g,1\rangle $ and $|e,0\rangle $ emerges. Thus EIT will occur in this composed system, and the state $|e,0\rangle $, $|g,1\rangle $ and $|g,0\rangle $ correspond to the three energy levels in of $\Lambda $ scheme of atomic EIT from highest to lowest \cite{[25]} as shown in Fig. 4.
\begin{figure}[tbph]
\centering\includegraphics[width=11cm]{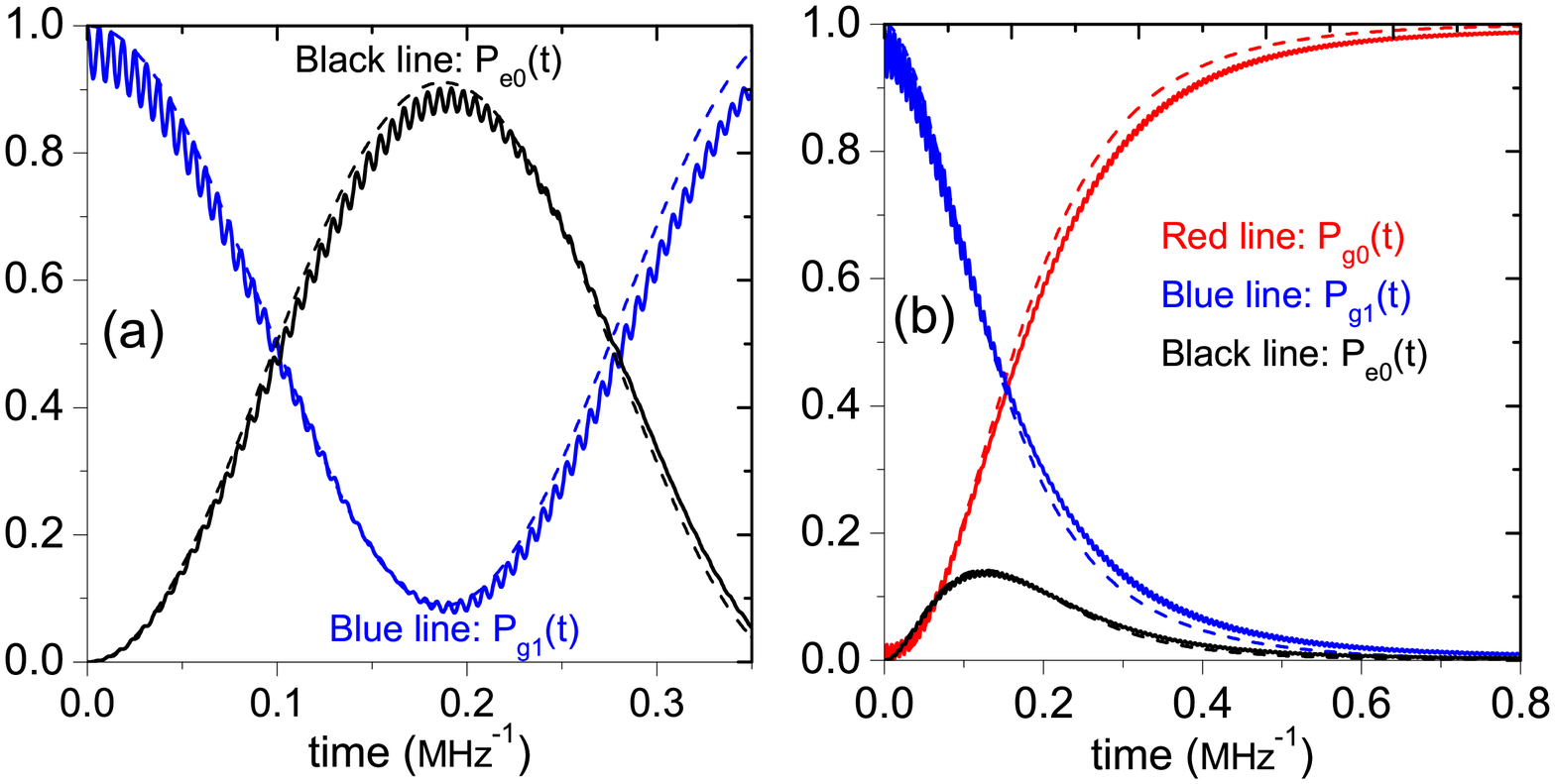}
\caption{Comparison of the populations based on the numerical results of the master equation (11) between $H_{eff}$ (dashed lines) and $H_{sys}^{^{\prime}}$ (solid lines). Here we set $\Delta =\protect\omega =1$ GHz$,$ $g=80$MHz and $\Omega _{pu}=50$MHz. For (a), we consider no decay channels, ie, $\Gamma_{d}=\Gamma _{f}=\protect\gamma =0$. For (b), we set the the decay strength $\Gamma _{d}/2=\Gamma _{f}=30$MHz and $\protect\gamma =0.1$MHz.}
\label{fig3}
\end{figure}

In fact we can make some approximations and reduce the expression of $\chi^{1}(\omega _{pr})$ in Eq. (7) to a more simple form by focusing on the
regime $\delta $ around $\omega $, which is similar to the resoled-sideband limit of optomechanical induced transparency in Ref. \cite{[8]}. Under the condition the pump-exciton detuning $\Delta $ is of much larger than pump strength $\Omega _{pu}$, we find that the flux qubit tends to stay in its ground state with assuming that the system decay to a vacuum reservoir, that is $\bar{\sigma}_{z}^{0}\simeq -1$ (This also can be evidenced in Fig. 3(b): the steady state of the system is $|g,0\rangle )$. Using another strength relation \{ $\Delta \simeq \omega \}\gg \{\Omega_{pu},g,\Gamma _{f}\}\gg \gamma $ and neglecting higher oder terms in the regime of $\delta \simeq \omega ,$ $\chi ^{1}(\omega _{pr})$ in equation (7) can be simplified as following
\begin{equation}
\chi _{eff}^{1}(\omega _{pr})=\frac{i\Gamma _{d}}{\Gamma _{f}-i[\delta-(\Delta +\frac{4g^{2}}{\omega }+\frac{2\Omega _{pu}^{2}}{\Delta })]+\frac{\Omega _{c}^{2}}{\frac{\gamma }{2}-i(\delta -\omega )}},
\end{equation}
where $\Omega _{c}=\frac{2g\Omega _{pu}}{\Delta }$ is the effective driving strength between dressed states $|g,1\rangle $ and $|e,0\rangle $, which is the same with the coupling rate in Eq. (10). The effective linear susceptibility $\chi_{eff} ^{1}(\omega _{pr})$ expressed in above equation have the same form with that of the atomic EIT in Refs. \cite{[5],[26]}. The corresponding energy level scheme is shown in Fig. 4. The energy shift $\Delta _{s}=\frac{4g^{2}}{\omega }+\frac{2\Omega _{pu}^{2}}{\Delta }$ between the states $|e,0\rangle $ and $|g,0\rangle $ are induced by the  \textit{LC} resonator (coupling terms $g(b^{\dagger }+b)\sigma _{z}\simeq $ $g\bar{X}_{0}\sigma _{z}\simeq $ $\frac{1}{2}(\frac{4g^{2}}{\omega })\sigma_{z})$ and the pump field ($\frac{2\Omega _{pu}^{2}}{\Delta }$, the first term demonstrated in Eq. (10)) collectively. Under the condition $\omega_{pu}=\upsilon -\omega $ ($\Delta =\omega )$, which equals that the second order pump field drives at rate $\frac{2g\Omega _{pu}}{\Delta }$ with detuning $\Delta _{s}$ (as shown in Fig. 4(a)), we plot the image and real part of $\chi ^{1}(\omega _{pr})$ and $\chi _{eff}^{1}(\omega _{pr})$ in Figs. 5(a) and (b), respectively. From the figure we can find $\chi^{1}(\omega _{pr})$ and $\chi _{eff}^{1}(\omega _{pr})$ match well around $\delta \simeq \omega $ under the adopted approximations.
\begin{figure}[tbph]
\centering\includegraphics[width=8cm]{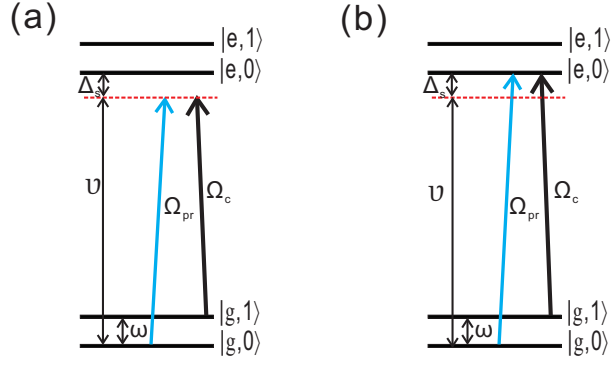}
\caption{Energy level schemes of the dressed states of the composed system. (a) shows the coherent driving between $|e,0\rangle $ and $|g,0\rangle $ at rate $\Omega _{c}=\frac{2g\Omega _{pu}}{\Delta }$ with detuning $\Delta _{s}=\frac{4g^{2}}{\protect\omega }+\frac{2\Omega _{pu}^{2}}{\Delta }$
(corresponding to the case $\Delta =\protect\omega )$, while (b) is associated to the case that $\Delta =\protect\omega -\Delta _{s}$, that is, the driving is on resonance.}
\label{fig4}
\end{figure}

If we set the pump-exciton detuning $\Delta =\omega -\Delta _{s}$ ($\Delta \simeq \omega $ is still valid since $\Delta _{s}\ll \omega )$ to compensate
the energy shift $\Delta _{s}$, and making sure the driving is resonant as shown in Fig. 4(b), Eq. (12) will reduce to a new simple form $\chi_{eff}^{1}(\omega _{pr})=\frac{i\Gamma _{f}}{\Gamma _{f}-i(\delta -\omega)+\Omega _{c}^{2}/[\frac{\gamma }{2}-i(\delta -\omega )]}$, which is the same with that of the atomic EIT occurs with a resonant driving field in Ref. \cite{[26]}. Figs. 5(c) and (d) show that $\chi _{eff}^{1}(\omega_{pr}) $ and $\chi ^{1}(\omega _{pr})$ match well and the transparency window is symmetrical, which proves the validation of our approximations. All these results indicate that the EIT in this composed system is similar to the atomic EIT. Moreover the differences are: in this EIT system the second order transfer between the dressed states serves as the pump field rather than a direct strong semiclassical field, and the energy shifts caused by coupling with the $LC$ resonator and driving field will affect the transparency window significantly.

According to the above discussions, it can be clearly found that our proposal provides an efficient method to realize EIT of currents in microwave lines with superconducting devices. The position of transparency window can be changed by modulating the applied flux $\Phi _{x}$ though the SQUID loop conveniently, so our proposal can work as a tunable EIT device for microwave field.
\begin{figure}[tbph]
\centering\includegraphics[width=10cm]{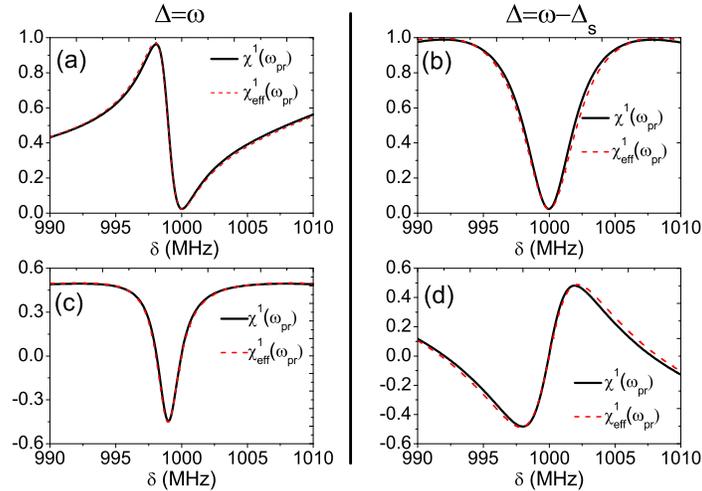}
\caption{(a) and (b) ((c) and (d)) show the Im$\protect\chi ^{1}(\protect \omega _{pr})$ (Re$\protect\chi ^{1}(\protect\omega _{pr}))$ as a function
of $\protect\delta $ under the condition $\Delta =\protect\omega $ (left panel) and $\Delta =\protect\omega -$ $\Delta _{s}$ (right panel). The red
dashed lines are determined by $\protect\chi _{eff}^{1}(\protect\omega_{pr}) $ while the black solid lines are determined by $\protect\chi ^{1}(\protect\omega _{pr})$.}
\label{fig5}
\end{figure}
\section{Conclusion}

In conclusion, we have proposed a composed EIT system in microwave regime using a superconducting flux qubit coupling with a $LC$ resonator. By obtaining the effective Hamiltonian of the system, we found that EIT occurs as a second-order coupling effect among the dressed states of the flux qubit and the $LC$ resonator. We showed some similarities and differences of the composed EIT comparing with the conventional atomic EIT, and interpret this phenomenon to some extent. Because the manufacturing technology for the superconducting devices is mature \cite{[27]}, it will provides us a convenient way to observe the EIT phenomenon in an artificial system. Since the close relations between EIT and slow light, our proposal may also work well as a slow light device integrated on chips.

\section*{Acknowledgments}

This work is supported by the Natural Science Foundation of China under grant No. 11174233.

\end{document}